\documentclass[pra,twocolumn, showpacs,nofootinbib]{revtex4}
\usepackage{epsf}
\usepackage{graphicx}
\usepackage{dcolumn}
\usepackage{epstopdf}

\begin{document}

\title{Proposed magneto-electrostatic ring trap for neutral atoms}
\author{Asa Hopkins}
\email{asa@caltech.edu}
\author{Benjamin Lev}
\author{Hideo Mabuchi}
\affiliation{Norman Bridge Laboratory of Physics 12-33, 
             California Institute of Technology, Pasadena, CA 91125} 
\date{\today}

\begin{abstract}
We propose a novel trap for confining cold neutral atoms in a microscopic ring using a magneto-electrostatic potential.  The trapping potential is derived from a combination of a repulsive magnetic field from a hard drive atom mirror and the attractive potential produced by a charged disk patterned on the hard drive surface.  We calculate a trap frequency of [29.7, 42.6, 62.8] kHz and a depth of [16.1, 21.8, 21.8] MHz for [$^{133}$Cs, $^{87}$Rb, $^{40}$K], and discuss a simple loading scheme and a method for fabrication.  This device provides a one-dimensional potential in a ring geometry that may be of interest to the study of trapped quantum degenerate one-dimensional gases.
\end{abstract}


\maketitle


Traps are a standard tool for the study and manipulation of cold neutral atoms, allowing the investigation of fundamental quantum dynamics as well as providing a basis for quantum information processing.  The manipulation of trapped atoms on ``atom chips" allows the implementation of many different atom optics elements for trapping, waveguiding, interferometry, etc.~\cite{JakobRev,SchmiedRev}.  Most atom chips use micron-sized current-carrying wires to generate the magnetic trapping fields.  We propose to construct a magneto-electrostatic ring trap, consisting of a hard drive atom mirror that provides a repulsive force on low-field seeking atoms~\cite{Lev03} and electric pads that attract polarizable atoms via the Stark effect~\cite{SchmiedElectric,Hinds,Krueger03}.  Schmiedmayer and Hinds and Hughes have proposed a range of such traps, including large-area two-dimensional traps, wire-based waveguides, and quantum-dot-like single state traps.  Such traps could be used to construct beam splitters or to implement collisional quantum gates~\cite{Calarco}.  Here we propose a novel ring trap for cold neutral atoms constructed from a conducting disk placed above the atom mirror surface, which produces a trap with a deep ring potential around the edge of the disk.


Let us first examine the trapping potential from a charged conducting disk above a hard drive atom mirror.  The hard drive's sinusoidal pattern of magnetization results in a repulsive potential---for atoms in weak-field seeking states---in the form of a decaying exponential~\cite{Opat} 
\begin{equation}
U_{mag} =  m_F g_F \mu_B B_0 \exp [- 2 \pi z/ a].
\end{equation}
The amplitude, $B_0$, depends on the remnant magnetization of the mirror as well as the magnetic sublevel $m_F$ and Land\'{e} $g_F$-factor of the atomic ground state.  The decay length is proportional to the periodicity $a$ of the magnetization pattern.  A small externally applied magnetic field perpendicular to the magnetization of the hard disk eliminates zones of zero magnetic field which would allow Majorana spin-flip losses.  The atom's low velocity allows the spin adiabatically to follow the magnetic field and thus the trapping potential depends only on the field magnitude.

In order to create a trap, the repulsive force from the mirror is balanced by an attractive force due to the DC Stark effect.  The atomic potential due to an electric field is
\begin{equation}
U_{Stark} = -\frac{1}{2}\alpha \left|E\right|^2,
\end{equation}
where we assume that we are working with atoms such as cesium or rubidium which possess only a scalar polarizability in the ground state.  A charged conducting disk creates high electric fields near its edge, resulting in a strong short-range attractive potential.

The mirror is made out of an etched hard drive whose aluminum substrate is grounded.  The boundary conditions consist of a ground at the mirror surface, and a constant potential on the surface of the thin conducting disk which is placed a distance $d$, typically on the order of a micron, above the mirror.  The electric fields are calculated from the solution to the Poisson equation with these boundary conditions.  The combined atomic potential due to the charged disk and mirror creates a trap above the conducting disk, which is deepest near the edge of the disk.

\begin{figure}
\centering
\includegraphics[width=3in]{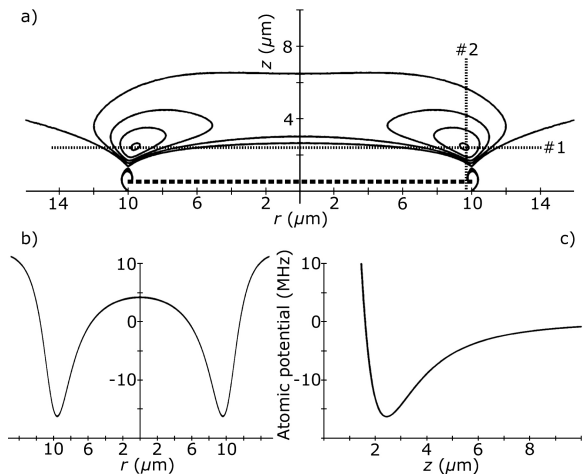}
\caption{\label{fig:potential} The atomic potential for cesium with 14.2 V on the disk.  a) A cross-section of the atomic potential in the plane containing the axis of the disk.  The contour lines are spaced 4 MHz apart.  The distance $r$ along a diameter of the disk and the distance $z$ above the disk are plotted on the horizontal and vertical axes, respectively.  b) The potential along slice \#1 in (a).  c)  The potential along slice \#2 in (a).}
\end{figure}

As an example, consider a conducting disk of radius 10 $\mu$m, placed $d=0.6$ $\mu$m above a hard drive atom mirror.  Let the hard drive have a field at its surface of 2 kG (a typical number for a commercial hard drive), and a periodicity of 3 $\mu$m in the magnetization.  The trapping potential for cesium in the $F = 3,$ $m_F =-3$ state near the edge of the disk has a depth of 16.1 MHz (770 $\mu$K) when the potential on the conducting disk is 14.2 V.  For $^{87}$Rb in the $F=2,$ $m_F = 2$ state, the trap has a depth of 21.8 MHz (1.05 mK) when 18.5 V is applied to the disk.  These two atomic states will be used in all examples for the remainder of the paper.  See Fig.~\ref{fig:potential} for the $^{133}$Cs potential.  The $^{87}$Rb potential looks qualitatively the same, with a slightly deeper minimum.  See Table ~\ref{table:cs} for trap parameters for a range of geometries for $^{133}$Cs and $^{87}$Rb, respectively.  For $^{40}$K, the optimal applied voltage is 4\% larger than that for the $^{87}$Rb trap, and trap frequencies scale up by a factor of $(m_{Rb}/m_{K})^{1/2} = 1.48$ relative to the $^{87}$Rb case.
\begin{table}
\centering
 \caption{\label{table:cs}Cesium 133 and rubidium 87 trap parameters for several disk radii $r$ and disk-hard drive separations $d$.}
\begin{ruledtabular}
\begin{tabular}{crlr@{.}lr@{.}lcc}
			&  				&			&  \multicolumn{2}{c}{ }	&    
\multicolumn{2}{r}{trap depth} 	& \multicolumn{2}{c}{trap frequencies} \\ 
$d$ ($\mu$m) 	& \multicolumn{2}{c}{$r$ ($\mu$m)} &  \multicolumn{2}{c}{V} 	&  \multicolumn{2}{c}{(MHz)}		& $\omega_{r}/2\pi$ (kHz) & $\omega_{\perp}/2\pi$ (kHz)\\
  \hline  
 \multicolumn{9}{l}{$^{133}$Cs} \\
\hline
0.6 	&5 		&&	13&3 	&~~~~17&0	&24.4	&44.1\\
0.6	&~~10 	&&	14&2 	&16&1		&29.7	&40.6\\
0.6	&20 		&&	14&8 	&15&4		&30.6	&37.0\\
1.0 	&5 		&&	9&4		&8&5		&18.0	&31.1\\
1.0	&10 		&&	10&2 	&8&2		&21.2	&28.0\\
1.0	&20 		&&	10&8 	&8&1		&22.1	&26.3\\
\hline
\multicolumn{9}{l}{$^{87}$Rb} \\
 \hline  
0.6	&5 		&&	17&3 	&22&9		&36.0	&63.4\\
0.6	&10 		&&	18&5 	&21&8		&42.6	&56.8\\
0.6	&20 		&&	19&2 	&20&6		&43.7	&52.7\\
1.0	&5 		&&	12&2 	&11&4		&25.7	&44.0\\
1.0	&10		&&	13&3 	&11&2		&30.8	&40.4\\
1.0	&20 		&&	14&0 	&10&7		&31.5	&37.5\\
\end{tabular}
\end{ruledtabular}
\end{table}


The curvature of the trap is large enough that the atom is confined in the Lamb-Dicke regime.  The Lamb-Dicke regime is defined as the regime in which $\eta = (E_{recoil}/E_{trap})^{1/2}< 1$.  For the parameters of Fig.~\ref{fig:potential}, the effective harmonic frequencies for [$^{133}$Cs, $^{87}$Rb, $^{40}$K] in the radial direction are [29.7, 42.6, 62.8] kHz, and [40.6, 56.8, 83.8] kHz in the direction perpendicular to the substrate.  We obtain a Lamb-Dicke parameter of $\eta \leq 0.26$ for $^{133}$Cs, $\eta \leq 0.30$ for $^{87}$Rb, and $\eta \leq 0.37$ for $^{40}$K.  Significantly higher trap frequencies are possible with the use of custom magnetic materials, which can have remnant magnetic fields of up to 2.4 T~\cite{Xue}.  For the same trap geometry as Fig.~\ref{fig:potential}, but using this custom magnetic material with a correspondingly higher applied voltage, the harmonic frequencies for $^{133}$Cs, for instance, are 103 kHz in the radial direction and 137 kHz in the perpendicular direction.  The higher remnant magnetic field also allows the disk to be placed further from the hard drive while maintaining significant trap depth.

\begin{figure}
\centering
\includegraphics[width=3in]{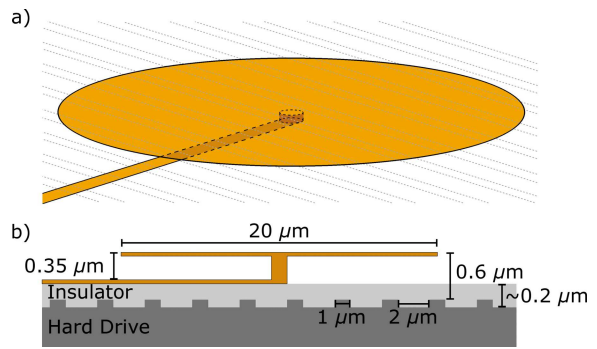}
\caption{\label{fig:disk}a) Schematic of the magneto-electrostatic ring trap drawn to scale.  The disk is 20 $\mu$m in diameter, with a 1 $\mu$m-wide lead connected via a central stem.  The dotted lines show the hard drive atom mirror's 2:1 etch pattern with a 3 $\mu$m periodicity.  b) Cross-section of the disk (with the vertical direction scaled up by a factor of 5), showing---from top to bottom---the disk, stem, lead, insulating layer, and etched hard drive.}
\end{figure}

A thin lead running along the hard drive surface may be used to connect the disk to a voltage source.  The maximum possible voltage on the disk is limited by the breakdown electric field of the dielectric material separating the lead from the conducting hard drive surface.  
In order to minimize the perturbation that the lead produces on the atomic potential from the disk, the lead should be as narrow as is practical ($\sim 1$ $\mu$m) and placed much closer to the hard drive surface than to the disk.  At this location, the repulsive force from the mirror is much stronger and no trap forms due to the charge on the lead.  In order to connect the lead to the disk, the disk is placed on a thin stem, with the lead connected to the bottom of the stem (see Fig.~\ref{fig:disk}).

Three dimensional solutions to the Poisson equation indicate that the effect from the lead on the trapping potential is minimized if the stem connecting the lead to the disk is located at the center of the disk.  For the previously used trap parameters and the lead placed $0.25$ $\mu$m above the hard drive surface ($0.35$ $\mu$m below the surface of the disk), the trap minimum for a $^{133}$Cs atom rises to $\sim 11.5$ MHz above the lead, which is a $\sim 30\%$ loss of trapping potential compared to the unperturbed trap.  The width of the perturbation is a few $\mu$m, slightly wider than the lead.  A shallower trap in which the electric pad is placed further from the mirror surface is perturbed less by the lead.  Use of custom magnetic materials would allow deeper traps to be constructed further from the lead, thereby minimizing the height of the perturbation.


We intend to fabricate the device as follows.  The hard drive atom mirror is etched in the manner described in Ref.~\cite{Lev03}, maintaining the 2:1 ratio of magnetization stripe spacing to minimize higher harmonics.  The stripe periodicity will be $\le3$ $\mu$m.  A deposition of a $\sim200$ nm thick insulating layer of silicon dioxide or silicon nitride is necessary to prevent shorting between the electric pads and the hard drive surface (see Fig.~\ref{fig:disk}).  This layer is thick enough to both support the voltage difference between the pads and underlying surface, and to help planarize the 100 nm deep corrugations of the etched hard drive.  The $\sim50$ nm tall, $\sim1$ $\mu$m wide gold leads are patterned on the insulator surface using standard photolithography and thermal evaporation of the adhesion metal and gold layers~\cite{Levfab,Madou}.  To create the stems, the surface is spin-coated with photoresist to a predetermined thickness to achieve optimal disk to atom mirror spacing.  Photolithography is again used to create vertical, cylindrical holes of 1 $\mu$m diameter in the photoresist located at the terminals of the gold leads.  The gold stems are electroplated from the gold leads through the cylindrical guide holes to the top of the photoresist.  A third photolithographic process and thermal evaporation patterns the 20 $\mu$m diameter gold disks attached to the tops of the stems.  Finally, the photoresist is removed using standard techniques, leaving behind the mushroom-like structures.  Field simulations show that to the percent level, leaving the photoresist under the disk does not disturb the electric field, and perturbations to the trap due to disk edge roughness or due to the hard drive trench corrugations are both negligible.  


The trap is conservative once the voltage is established, and the kinetic energy of the atoms must be lowered for them to stay in the trap.  A simple, but inefficient, method of loading this trap is to drop a cloud of cold atoms from a magneto-optical trap (MOT)---sub-doppler cooled to 10 $\mu$K---onto the device.  The atoms are captured by turning on the voltage on the electric pads as the atoms are passing through their classical turning point above the atom mirror.  Simulations indicate that this scheme can capture 1 to 2$\%$ of the dropped atoms.  The fraction is small because the voltage ramp must be quite fast ($\sim 2 \times 10^{-4}$ seconds) in order to remove enough energy from the atoms to trap them, while the atom cloud takes roughly $2 \times 10^{-2}$ seconds to pass through the trapping volume.  This scheme has many different parameters over which loading can be optimized, including the initial position, size and density of the MOT before it is dropped, and the shape and speed of the voltage ramp.  Ramping up the voltage on the conducting disk is the simplest scheme for trapping the atoms, but it is possible that another procedure, involving atomic transitions or other degrees of freedom in the system, could be more effective and is currently being investigated.

Given this loading efficiency, a 10 micron-radius ring trap will capture roughly 30-50 atoms from a dropped cloud of 10$^7$ atoms and temperature 10 $\mu$K.  In order to capture more atoms, disks can be arranged in an array covering a larger surface area.  The volume of the trap deeper than 200 $\mu$K is 1 to $2 \times 10^{-9}$ $\rm{cm}^3$.  Simulations indicate that these traps can be placed roughly 20 $\mu$m apart without significantly disturbing each other.  Therefore, roughly $20 \%$ of the surface can be covered with the traps.  Combining the loading efficiency with this surface coverage, roughly a few $10^3$ atoms can be trapped.  The leads can be routed though spaces between the disks with either a separate lead for each disk or a shared network of leads.

Several undesired effects, such as heating, fragmentation of Bose-Einstein condensates, and a reduction of trap lifetimes, have been detected in microtrap experiments involving atoms near room-temperature surfaces.  The trap proposed here is insusceptible to heating due to technical noise on the currents in the microwires and to the fragmentation problems caused by the spatial variation of these currents~\cite{Fortagh02b,Leanhardt03}.  However, the trap remains susceptible to atom loss due to spin flips induced by magnetic field fluctuations from thermal currents in the metal forming the electric pads, as detected in several experiments~\cite{Jones03,Cornell03,Vladan03}.  Surface effects in this system will most closely resemble those in Y. Lin {\it et al.}, wherein the skin depth for the transition frequency between trapped and untrapped magnetic sublevels of the atoms is much larger than both the distance of the atoms from the metal surface and the thickness of the metal conductor.  As reported in Y. Lin  {\it et al.}, at a distance of 2 $\mu$m this Johnson noise limits the lifetime of $^{87}$Rb atoms above a 2 $\mu$m thick copper conductor to a few 100 ms---ample time for detecting atoms in the ring trap.  The metal film used for the electric disk pad in the ring trap will be ten to a hundred times thinner than that used for the above experiment, and we expect this to further minimize the trap's loss rate~\cite{Henkel03,Vladan03}. 


There are several possibilities for minimizing or eliminating the perturbation due to the lead, or tuning it to be of a particular height, other than simply adjusting the trap geometry.  The most versatile possibility is to add an additional photolithography step to insert another electric pad directly above the lead, separated by a thin insulation layer.  The voltage applied to this separate pad can be used to compensate for the effects of the lead.  In particular, the voltage on a pad the same width as the lead and placed 100 nm above it could be tuned to completely eliminate the perturbation (to within the percent-level accuracy of our calculations) or turn it into a dip rather than a bump.  Complete elimination of any perturbation is possible by expanding such a pad to cover the entire surface, with a hole to allow the stem to reach from the lead to the disk.  Another possibility is to charge the disk not with a lead but with an intrachamber electron beam.  Such a system would be hard to charge and discharge quickly, requiring a loading scheme that does not require a rapid change to the charge distribution.  

Several future improvements or extensions of this trapping concept are possible.  For example, the decoherence effects due to the proximity of a conductor could be mitigated with the use of a dielectric magnetic film in place of the hard drive, and dielectric pads charged via an electron beam in place of the conducting disks.  In addition, disks, rings, wires, and other shapes could be used to trap and manipulate the atoms just above the surface, and voltages adjusted to shift the atoms from one potential into another.  Integration of these traps with magnetic microtraps based on current-carrying wires on the surface is also possible.  Small single-atom traps with additional electrostatic pads to control the barrier heights could enable a system for quantum logic gates~\cite{Calarco}.

In the past several years, there has been much experimental and theoretical interest in trapped one-dimensional (1D) quantum degenerate gases (see references~\cite{Esslinger03,Porto03,Olshanii98,Walraven00,Jakob04} and the citations within).  Trapped 1D gases require $k_{B}T, \mu\ll \hbar\omega_{\perp}$, where $T$ is the temperature, $\mu$ is the chemical potential, and $\omega_{\perp}$ is the transverse trapping frequency.  Various regimes of quantum degeneracy---of which a 1D gas of impenetrable bosons, the Tonks-Girardeau (TG) regime, is of particular interest---can be explored by changing the density of trapped atoms or by modifying the interactions between atoms via Feshbach resonances.  In the latter case, a magnetic bias field for adjusting the s-wave scattering length, $a$, can be added parallel to the magnetization stripes of the atom mirror without affecting the potential of the magneto-electrostatic ring trap.  With respect to $^{87}$Rb, a common alkali used for BEC, $k_{B}T/\hbar\omega_{\perp}$ is smaller than 0.05 for temperatures below 100 nK.  The TG regime requires that the mean interparticle separation, $1/n$, be much larger than correlation length, $l_c=(\hbar/2mn\omega_{\perp}a)^{1/2}$, where $m$ is the atom's mass, $n=N/L$ is the number density~\cite{Walraven00}.  This constraint limits the number of $^{87}$Rb atoms in the ring trap to $N\ll 2m\omega_{\perp}aL/\hbar=250$ atoms for a device of circumference $L=2\pi\cdot20$ $\mu$m, $\omega_{\perp}=2\pi\cdot40$ kHz, and an $a$ unmodified by Feshbach resonances (the field at the trap minimum is $\sim12$ G).  Overcoming the challenge of detecting so few atoms may be possible through the incorporation of microwire traps~\cite{Jakob04}.

The ring geometry adds a unique element to the many-body physics of the 1D trap.  Josephson effects in trapped BECs have been investigated theoretically for the case of a double well (see reference~\cite{Williams01} and citations within) and investigated experimentally in an optical standing wave~\cite{Kasevich98}.  A BEC in this magneto-electrostatic ring trap system with interspersed Josephson junctions formed from the addition of micron-sized perturbations to the trapping potential---such as those caused by wire leads, possibly tuned using additional pads---is reminiscent of superconducting electronic systems.  The ratio of the chemical potential to the perturbation barrier height can be adjusted with the trap parameters such as $d$, $r$, atom number, and disk potential, as well as the use of additional electric pads, to cause the perturbation to act as either an impenetrable wall, a tunnel junction, or a scattering center.  The utility of this 1-D ring trap is highlighted by recent proposals for using a BEC in a double ring to create a SQUID-like device for neutral atoms~\cite{Anderson} and for investigating quantum chaos in the system of the quantum kicked rotor~\cite{Raizen03}.

This magneto-electrostatic trap for cold neutral atoms---derived from balancing the repulsive force of an atom mirror with the attractive force from a charged disk---introduces a novel ring trapping geometry for cold neutral atoms.  Fabrication of this trap is straightforward, and an array of such traps can trap a significant number of atoms.  Furthermore, such a trap may allow the exploration of interesting many-body physics in a one-dimensional ring trap.  This device is an example of the rich potential for developing novel atom optical elements through the integration of a hard drive atom mirror, charged pads, and microwires.


The authors thank Axel Scherer for helpful discussions.  This work was supported by the Multidisciplinary University Research Initiative program under Grant No. DAAD19-00-1-0374.  A.H. acknowledges an NSF Graduate Research Fellowship.

\end{document}